\documentclass[
  reprint,
  aps,
  prb,
  superscriptaddress,
  amsmath,amssymb
]{revtex4-2}
\usepackage{xcolor}
\usepackage{bm}
\usepackage[squaren]{SIunits}
\usepackage{physics}
\usepackage{graphicx}
\usepackage[english]{babel}
\usepackage{orcidlink}
\usepackage[normalem]{ulem}

\usepackage{hyperref}
\hypersetup{colorlinks=true,linkcolor=blue,citecolor=blue,urlcolor=blue}

\begin{document}

\title{Wave Transport in Fourier Quasicrystals Revealed by Water Waves}

\author{Angélique Campaniello \orcidlink{0009-0006-5531-7031}}
\thanks{angelique.campaniello@espci.psl.eu} 
\affiliation{Institut Langevin, ESPCI Paris, PSL University, CNRS, Paris, France}

\author{Lior Alon  \orcidlink{orcid.org/0000-0002-0453-1189}}
\thanks{lioralon@mit.edu} 
\affiliation{School of Mathematics, Massachusetts Institute of Technology, Cambridge, MA 02139, USA}

\author{Rémi Carminati \orcidlink{0000-0002-5839-1820}}
\thanks{remi.carminati@espci.psl.eu} 
\affiliation{Institut Langevin, ESPCI Paris, PSL University, CNRS, Paris, France}
\affiliation{Institut d’Optique Graduate School, Paris-Saclay University, 91127 Palaiseau, France}

\author{Emmanuel Fort \orcidlink{orcid.org/0000-0003-2770-3753}}
\thanks{emmanuel.fort@espci.psl.eu} 
\affiliation{Institut Langevin, ESPCI Paris, PSL University, CNRS, Paris, France}

\author{Marcel Filoche \orcidlink{0000-0001-8637-3016}}
\thanks{marcel.filoche@espci.psl.eu} 
\affiliation{Institut Langevin, ESPCI Paris, PSL University, CNRS, Paris, France}

\date{\today}

\begin{abstract}
Fourier quasicrystals are aperiodic structures whose diffraction spectrum consists not of a dense set of Bragg peaks, as in ordinary quasicrystals, but of isolated ones scattered across a discrete, nonperiodic set. This sparse reciprocal-space structure should leave wave transport largely undisturbed except at a few selected wavevectors. We put this prediction to the test using surface water waves scattering off a two-dimensional Fourier quasicrystal. Full-field measurements reveal three distinct transport regimes as the incident wavevector increases: transparency, selective scattering, and strong scattering. By reconstructing the structure factor from the measured wavefields, we directly relate these regimes to the underlying reciprocal-space structure. Our results establish Fourier quasicrystals as a physical platform in which wave transport can be controlled through the organization of diffraction peaks in reciprocal space.
\end{abstract}

\maketitle

While periodic crystals exhibit discrete Bragg peaks arranged on reciprocal periodic lattices, all known two- or three-dimensional quasicrystals---displaying long-range order without translational symmetry---were characterized until recently by a dense pure-point structure factor, i.e., a dense set of Bragg peaks in reciprocal space. The discrete rotational symmetry emerges explicitly only when thresholding the intensity of these diffraction peaks above a finite value~\cite{Levine1984, Shechtman1984, Meyer1972, Lagarias2000,Moody2000}. In such structures, wave propagation has revealed a wide range of unconventional phenomena, including acoustic and photonic localization \cite{Macon1991, Kahyun2017}, disorder-enhanced transport~\cite{Levi2010}, matter-wave diffraction \cite{Viebahn2018}, directional wave propagation~\cite{Danilo2022}, diffraction signatures of topological order~\cite{Dareau2017}, and complex spectral organization~\cite{Vignolo2016}.

A distinct class of quasicrystals has recently emerged: \emph{Fourier quasicrystals} (FQCs), whose structure factor consists of \emph{isolated} Bragg peaks rather than a dense set~\cite{Kurasov2020, Alon2025, Lawton2025}. The most recent of these results, from one of the present coauthors, established the existence of FQCs in dimension higher than one and provided a systematic construction~\cite{Alon2025}. Unlike conventional quasicrystals, FQCs concentrate their diffracted intensity on a discrete, nonperiodic subset of reciprocal space, leaving extended regions free of any Bragg peak. This sparsity should strongly constrain the momentum transfers available to a scattered wave, driving a transition from transparency to selective scattering as the wavenumber increases, and eventually to strong quasi-isotropic scattering once enough Fourier components become accessible. Here we report the first experimental test of this picture, using surface water waves, and show that the predicted transport regimes emerge even in a finite sample and an absorbing medium.


Wave transport is governed by the structure factor~$S(\vb{q})$, where $\vb{q}=\vb{k}_s-\vb{k}_i$ is the scattering vector associated with the momentum transfer between incident and scattered waves with wavevectors~$\vb{k}_i$ and $\vb{k}_s$, respectively. For $N$ scatterers located at positions~$\vb{r}_j$, the structure factor is defined as
\begin{equation}
   S(\vb{q}) = \frac{1}{N} \abs{ \sum_j e^{-i\,\vb{q} \cdot \vb{r}_j} }^2 \,.
\end{equation}
Elastic scattering is allowed only for wavevectors belonging to the support of $S(\vb{q})$~\cite{Sheng_introduction_2006, Sebbah_waves_2001, Akkermans_Montambaux_2007, Carminati_principles_2021, Vynck23, Wiersma13, Yu20, Cao_harnessing_2022}.

\begin{figure}[!h]
    \centering
    \includegraphics[width=\linewidth]{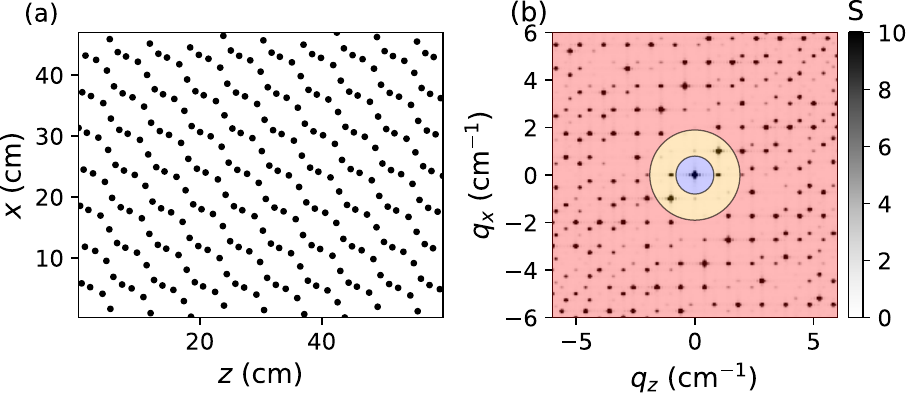}
\caption{\textbf{Fourier quasicrystal}. (a)~Real-space scatterer arrangement. (b)~Corresponding structure factor $S(\vb{q})$ (dimensionless quantity). A depleted region is observed around $\vb{q} = \vb{0}$, where $S(\vb{q}) \approx 0$, followed by an increasing density of peaks at larger $\abs{\vb{q}}$.}
    \label{fig:structure factor}
\end{figure}

The FQC used in this work consists of subwavelength scatterers placed on a two-dimensional point pattern $(x,z)$ generated following the general construction introduced in Refs.~\cite{Kurasov2020, Alon2025}. In our case, each point $(x,z)$ solves the following system:
\begin{align}
\begin{cases}
 \cos(\frac{Z_1-Z_2}{2}) - \cos(\frac{Z_1+Z_2}{2}) + 2 \sin(\frac{Z_1+Z_2}{2}) = 0 \\\\
 -\cos(\frac{Z_1-Z_3}{2}) + \cos(\frac{Z_1+Z_3}{2}) + \sin(\frac{Z_1-Z_3}{2}) = 0
\end{cases}
\end{align}
with $(Z_1,Z_2,Z_3) = z \, \omega_1 + x \, \omega_2 $, $\omega_1=\left(1,0,-\sqrt{2}\right)$ and $\omega_2=\left(0,1,\sqrt{3}\right)$.
The resulting point pattern is nonperiodic yet possesses a diffraction spectrum composed of isolated Bragg peaks. Its real-space arrangement and corresponding structure factor are shown in Fig.~\ref{fig:structure factor}. In particular, we observe that the structure factor exhibits a depleted region around~$\vb{q}=\vb{0}$, followed by isolated Bragg peaks whose density increases with distance from the origin. This shows that FQCs, in stark contrast with usual quasicrystals built from aperiodic tilings such as Penrose tilings, are stealthy hyperuniform media~\cite{torquato_hyperuniform_2018, Leseur16}.

\begin{figure}[h]
    \centering
    \includegraphics[width=\linewidth]{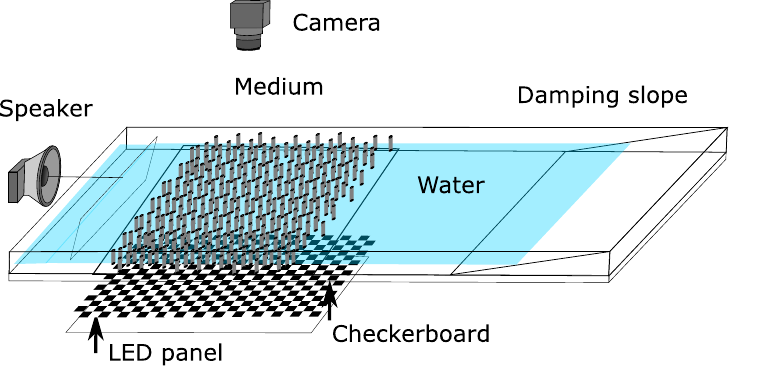}
\caption{\textbf{Schematic of the experimental setup}. A loudspeaker-driven paddle generates monochromatic surface waves in a \unit{150 \times 60}{cm} tank. Surface elevation is reconstructed from the deformation of a checkerboard beneath the transparent bottom. A sloped beach suppresses reflections.}
    \label{fig:experimental set-up}
\end{figure}

The diffraction properties of this structure (or a finite version of it) are tested using surface water waves in the \unit{3.5-9}{Hz}~frequency range. Experiments are carried out in a water tank, where surface waves are generated by a horizontally oscillating paddle, producing a quasi-plane incident wave~\cite{Campaniello_PNAS}. The scattering medium, of dimensions \unit{45\times60}{cm}, consists of cylindrical scatterers of diameter $d = \unit{1}{cm}$ arranged in a FQC configuration with $N=287$~scatterers, corresponding to a surface density $\rho \simeq \unit{0.1}{cm^{-2}}$ (Fig.~\ref{fig:experimental set-up}). The wavefield is reconstructed using Fourier checkerboard demodulation~\cite{Wildeman18}, yielding the complex surface elevation. A lock-in analysis at the driving frequency is subsequently applied to extract the amplitude and phase fields. Further experimental details are provided in Appendix~\ref{Appendix:setup}.


Plane waves are sent through the structure at different frequencies, here $4.1$, $5.4$, $\unit{6.2}{Hz}$, respectively, corresponding to wavelengths in the several centimeter range (the dispersion relation is given in Appendix~\ref{Appendix:setup}). Figure~\ref{fig:fields} displays the wavefield amplitudes (top row), the phase fields (middle row), and the spatial Fourier transform analysis (bottom row), respectively. Prior to computing the spatial Fourier transforms, the measured fields were multiplied by a two-dimensional Hann window to reduce spectral leakage arising from the finite observation domain. The three columns of Fig.~\ref{fig:fields} correspond to the three aforementioned frequencies. In all cases, the wave amplitude decreases along the propagation direction due to viscous dissipation, with an absorption length of about~\unit{20}{cm} (Fig.~\ref{fig:fields}a-c). The measured wavefields also reveal a clear increase of wave scattering with increasing frequency. Yet, a much more interesting structure emerges when looking at the phase field.
\begin{figure*}[ht]
    \centering
    \includegraphics[width=0.6\linewidth]{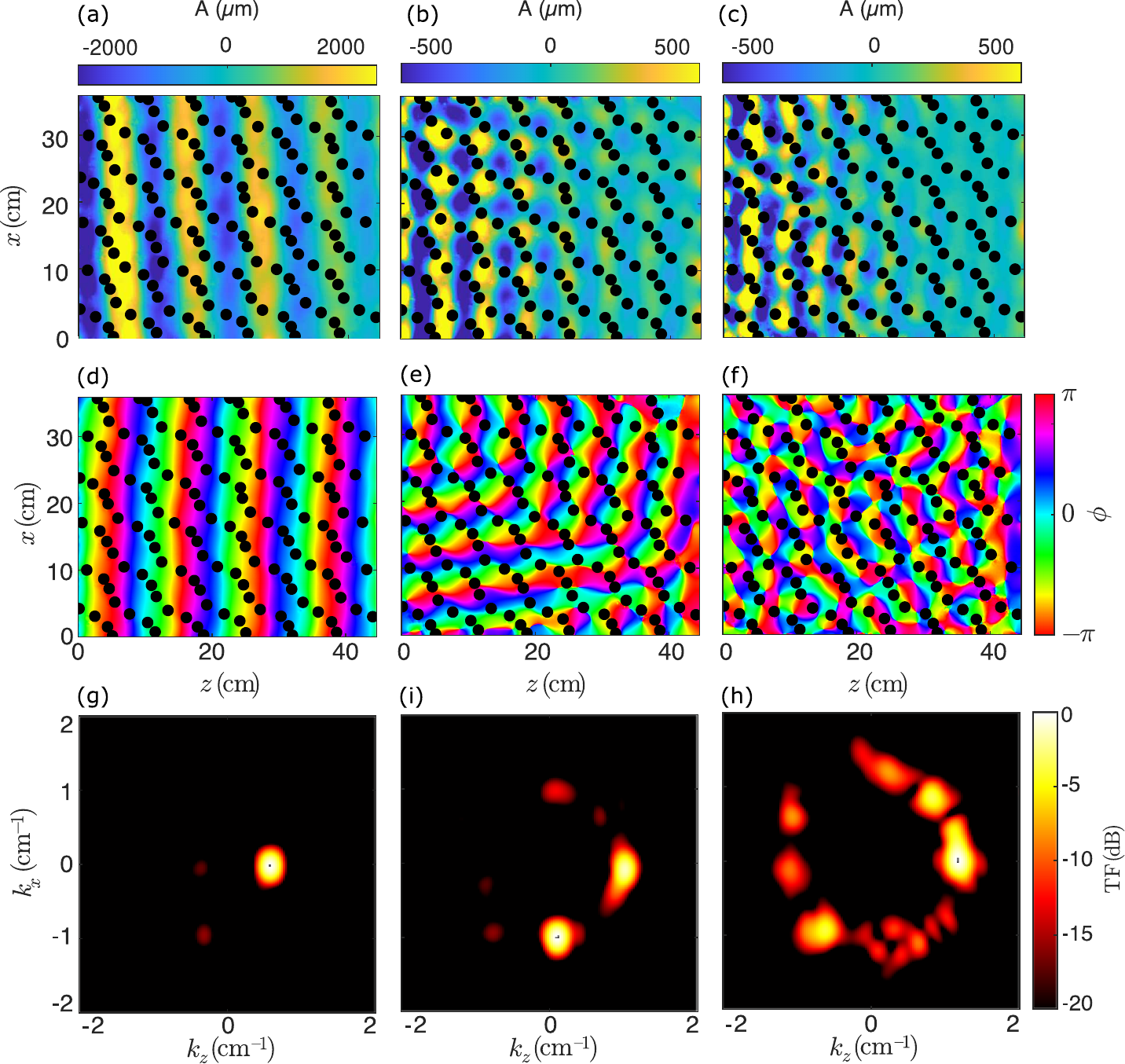}
\caption{\textbf{Measured surface elevations, phases, and spatial Fourier transforms}. (a–c)~Real surface-elevation maps at $4.1$, $5.4$, and $\unit{6.2}{Hz}$, showing non-scattering, selective scattering by a Fourier peak, and strong scattering, respectively. The incident wave propagates from left to right. Black dots indicate scatterer positions. The colormaps have been slightly saturated to enhance wavefront visibility. (d–f)~Corresponding phase maps. (g–i)~Hann-windowed spatial Fourier transforms Additional peaks emerge with increasing wavevector, revealing the onset of scattering.}
    \label{fig:fields}
\end{figure*}

At low frequency, the phase pattern remains essentially undistorted across the medium (Fig.~\ref{fig:fields}d), indicating that the incident plane wave propagates through the structure without scattering. Consistently, the corresponding Fourier spectrum is dominated by the incident wavevector (Fig.~\ref{fig:fields}g). At intermediate frequency, the phase fronts become tilted and are redirected toward the lower-right part of the medium (Fig.~\ref{fig:fields}e). This directional deflection is accompanied by the appearance of an additional peak in the Fourier spectrum (Fig.~\ref{fig:fields}h), a behavior similar to diffraction by a crystal. At higher frequency, the phase field becomes strongly distorted, displaying a speckle-like structure (Fig.~\ref{fig:fields}f), reflecting the onset of multichannel scattering. Correspondingly, numerous peaks emerge in the Fourier spectrum (Fig.~\ref{fig:fields}i), leading to a quasi-isotropic pattern, analogous to diffuse propagation through a disordered medium. Notably, these peaks are nevertheless all located on a circle centered on $\vb{q} = \vb{0}$, signature of elastic scattering. This progressive transition from transparency to directional scattering, and finally to strong multichannel scattering is consistent with the reciprocal-space organization of the FQC, whose structure factor suppresses scattering near $\vb{q} = \vb{0}$ while supporting a number of discrete scattering vectors that is increasing at larger $|\vb{q}|$.


Conversely, one can use these experiments to reconstruct the structure factor experienced by the incoming plane waves. The structure factor is reconstructed from the measured wavefields by processing the spatial Fourier transforms at each excitation frequency. For each transform, an annular region of width \unit{0.1}{cm^{-1}}, of the order of the experiment's resolution, corresponding to the elastic scattering shell is selected and recentered by subtracting the incident wavevector, thereby expressing the data in terms of the scattering vector $\vb{q}$. This procedure is repeated over the full frequency range and for the four incident-wave orientations. For each orientation, the recentered annuli are stacked to reconstruct a reciprocal-space disk. Since different regions of a given disk are sampled a different number of times, the stacked intensity is normalized by the number of contributions accumulated in each pixel. Details of the procedure are provided in Appendix~\ref{Appendix:reconstruction}.

\begin{figure}[h]
    \centering
    \includegraphics[width=\linewidth]{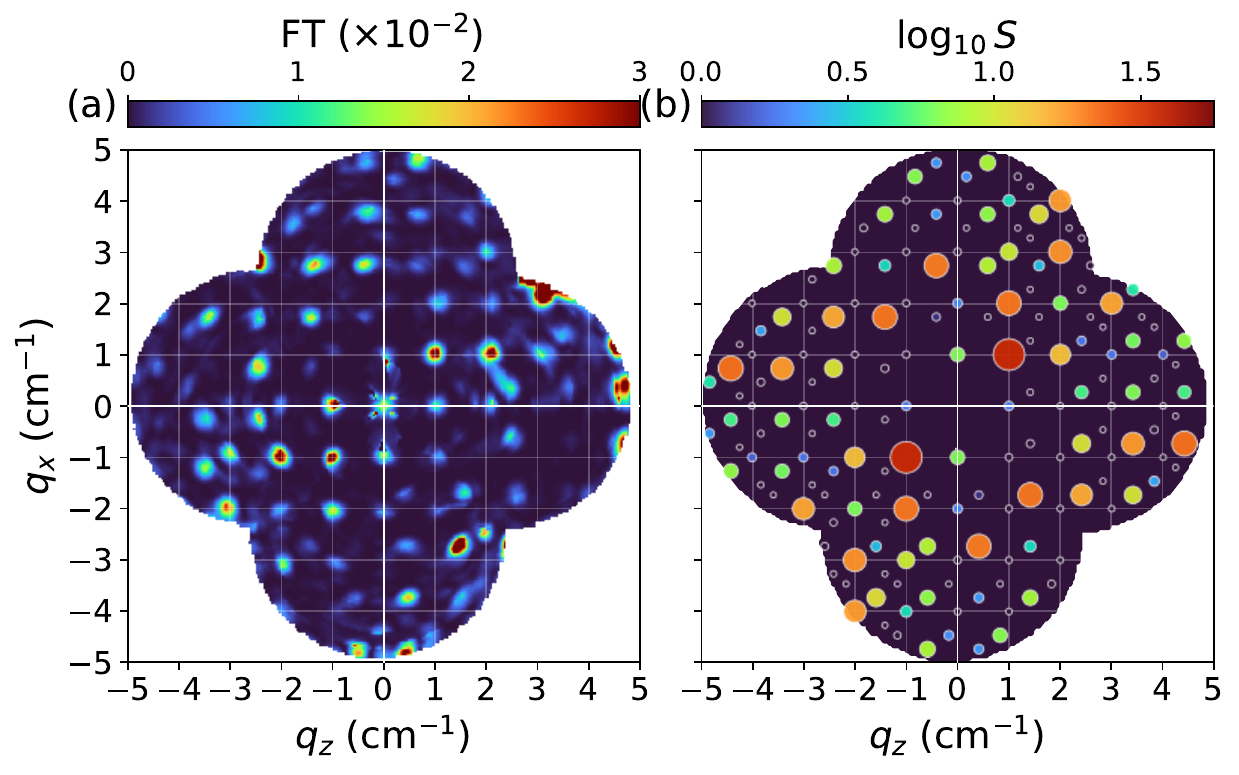}
\caption{ \textbf{Reconstruction of the FQC structure factor from measurements}. (a)~Reciprocal-space intensity obtained by stacking the elastic-scattering shells measured between $3.5$ and \unit{9}{Hz} for four incident-wave orientations. (b)~Theoretical structure factor restricted to the experimentally accessible reciprocal-space region, with circle sizes and colors proportional to the peak intensities. The dominant peaks observed experimentally are well captured by the theoretical prediction. }
    \label{fig:reconstructed structure factor 2}
\end{figure}

The four reconstructed disks are then combined to obtain the full reciprocal-space map. In regions where disks overlap, the minimum intensity value is retained, which reduces reconstruction artifacts. Small angular corrections, not exceeding $2^\circ$, are also applied to each orientation to compensate for slight experimental misalignment. A small Gaussian smoothing is applied during this procedure to improve the visibility and continuity of the reconstructed peaks. The reconstructed distribution is shown in Fig.~\ref{fig:reconstructed structure factor 2}a. For comparison, Fig.~\ref{fig:reconstructed structure factor 2}b displays the theoretical structure factor of the infinite FQC restricted to the experimentally accessible region of reciprocal space. Only peaks lying within the experimentally accessible reciprocal-space region were retained for comparison with the reconstructed spectrum.


Overall, the reconstructed structure factor is in very good agreement with the theoretical prediction (Fig.~\ref{fig:reconstructed structure factor 2}b). In particular, the dominant Fourier peaks are recovered at the expected positions, demonstrating that scattering of water waves captures the reciprocal-space organization of the Fourier quasicrystal. This agreement holds despite the finite size of the experimental medium and the presence of absorption, showing that our reconstruction successfully recovers the reciprocal-space structure of the underlying infinite scattering system. Minor discrepancies still remain, including slight shifts in peak positions and the absence of some smaller peaks. In addition, the increasing viscous attenuation at higher frequencies degrades the reconstruction quality, leading to broadened features and a reduced visibility of peaks associated with larger scattering vectors. Nevertheless, the dominant features of the structure factor are accurately reproduced up to at least $q \simeq \unit{4}{cm^{-1}}$.

These results show that, despite their limitations due to viscous absorption and finite size, the experiments clearly probe the three regimes targeted in this work, i.e., transparency, Bragg diffraction, and strong multichannel scattering.

This transition between different regimes observed here, that combines features of both hyperuniform, crystalline, and fully ordered structures, appears to be specific to FQCs. As in stealthy hyperuniform systems, scattering is suppressed over a finite region around $\vb{q} = \vb{0}$. However, when scattering becomes allowed, it occurs through a discrete set of reciprocal-space peaks rather than through a continuous spectrum. The distinctive feature of FQCs is therefore the possibility of engineering the radial distribution of these peaks, and hence the wavevector dependence of scattering. Our results provide an experimental demonstration of this mechanism for water waves and establish FQCs as a versatile platform for reciprocal-space engineering of wave transport.


In conclusion, we have demonstrated the first experimental realization of a Fourier quasicrystal using surface water waves. Its scattering behavior departs qualitatively from that of conventional quasicrystals: hyperuniform at low frequency, fully diffusive at high frequency, with a controllable crossover in between. These properties, confirmed by our full-field measurements, establish water-wave Fourier quasicrystals as a versatile, tabletop platform for reciprocal-space engineering of wave transport, with implications extending to acoustics, photonics, and beyond.

\begin{acknowledgments}
This work has received support under the program ``Investissements d’Avenir'' launched by the French Government. M. Filoche and A. Campaniello are supported by the project Localization of Waves of the Simons Foundation (Grant No.~1027116, M.F.). E. Fort is supported by the AXA Research Fund. L. Alon is supported by the project Localization of Waves of the Simons Foundation (Grant No.~1027116, D.J.)
\end{acknowledgments}

\appendix

\section{Experimental setup}
\label{Appendix:setup}

\subsection*{Water tank and wave excitation}

The experiments are performed in a rectangular wave tank of dimensions $\unit{150}{cm} \times \unit{60}{cm}$, filled with water of depth~\unit{7}{cm}. Surface waves are generated by a wave maker consisting of a \unit{40}{cm}-wide vertical paddle actuated by a loudspeaker. The paddle is driven by a sinusoidal signal produced by a function generator (Rigol DG4162) and amplified using an audio amplifier (König AMP 4800).

Measurements are conducted over the frequency range \unit{3.5-9}{Hz}. Because the paddle width remains significantly larger than the wavelength, the incident field can be accurately described as a plane wave over the measurement region. To minimize reflections from the tank boundaries, a sloping absorbing beach is installed at the opposite end of the basin.

In the linear wave regime, assuming the fluid is curl-free and incompressible, the surface wave can be described as a 2D scalar wave satisfying the Helmholtz equation. The waves obey the gravity-capillary dispersion relation \cite{Landau, Campaniello_PNAS} :
\begin{equation}
    \omega^2= \left(gk+\frac{\gamma}{\rho_w}k^3\right)\tanh(kh),
\end{equation}
where $\omega$ is the frequency, $k$ the wavenumber, $h$ the water depth, $\gamma$ the surface tension, $\rho_w$ the water density. Wave propagation is also affected by dissipation, including bulk viscosity, friction on the tank and surface contamination, resulting in an exponential decay of the wave amplitude. 

\subsection*{Fabrication of the Fourier quasicrystal}

The Fourier-quasicrystal sample consists of $N=287$~cylindrical PMMA scatterers (\unit{1}{cm} diameter, \unit{8}{cm} height) inserted into a rectangular PMMA plate measuring $\unit{45}{cm} \times \unit{60}{cm}$ and \unit{1}{cm} thick. Circular holes matching the scatterer diameter are laser-cut into the plate to ensure accurate positioning of the cylinders. The plate thickness was chosen to provide sufficient mechanical rigidity and prevent bending of the support. The resulting scatterer density is $\rho = \unit{0.1}{cm^{-2}}$. Throughout the investigated frequency range, the cylinders behave as non-resonant subwavelength scatterers. The scatterer positions are determined from a FQC point pattern generated following the construction described in Refs.~\cite{Alon2025, Lawton2025}.

The PMMA plate can be rotated by $0^\circ$, $90^\circ$, $180^\circ$ or $270^\circ$, to explore different regions of the structure factor.

\subsection*{Wavefield measurement and reconstruction}

Wave fields are recorded by imaging the distortion of a checkerboard pattern located beneath the transparent tank bottom. The images are acquired with a Basler acA1300-200um camera fitted with a \unit{16}{mm}~F/1.4 lens and positioned approximately \unit{2}{m} above the water surface. Frame rates between $35$ and \unit{100}{fps} are used, ensuring a temporal sampling of at least ten images per oscillation period. For each measurement, a sequence of 200~frames is recorded, corresponding to roughly twenty wave periods.

\subsection*{Scatterer masking}

The regions occupied by the scatterers are excluded from the analysis before wavefield reconstruction. A binary mask is generated from a reference image by detecting the strong intensity gradients associated with the scatterer boundaries. The resulting mask is expanded by approximately five pixels to ensure complete removal of near-scatterer contributions. This mask is applied directly to the raw images before demodulation, reducing artefacts arising from the near field of the cylinders.

 \subsection*{Amplitude and phase extraction}

Following the wavefield reconstruction, a spatial high-pass filter is applied to suppress the low-spatial-frequency background introduced by the imaging and reconstruction procedure. The complex wave field at the excitation frequency is then extracted using a lock-in approach, implemented by isolating the corresponding temporal Fourier component of the measured signal. This procedure yields both the amplitude and phase distributions of the wave field.

The analysis is performed for both the total field, measured in the presence of the scatterers, and a reference incident field recorded in their absence. The scattered field is subsequently obtained by phase-aligning the reference field with the total field and subtracting the two complex wave fields.

\section{Experimental reconstruction of the structure factor}
\label{Appendix:reconstruction}

\begin{figure}[h]
    \centering
    \includegraphics[width=0.9\linewidth]{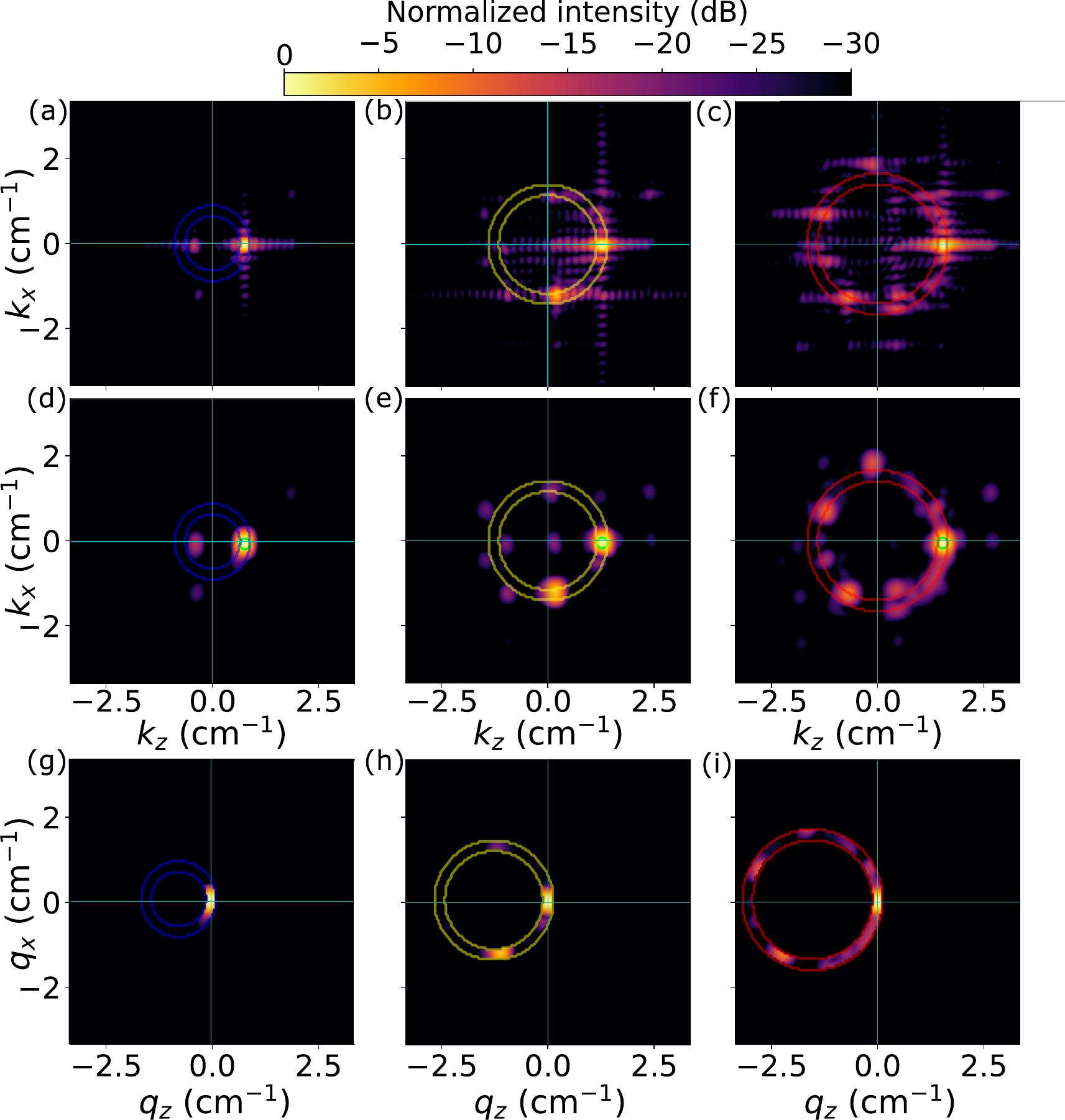}
    \caption{
\textbf{Extraction of the elastic Ewald circles in Fourier space for three excitation frequencies}.
The top row (a, b, c) shows the Fourier spectra of the measured wave field without spatial windowing, while the middle row (d, e, f) displays the same spectra after applying a Hann window to reduce edge effects. The peak associated with the incident wave is visible along the horizontal axis, and the corresponding elastic scattering circle, imposed by conservation of the wavevector magnitude, is indicated by a colored contour (blue: \unit{4.1}{Hz}, yellow: \unit{5.4}{Hz}, red: \unit{6.2}{Hz}). The bottom row (g, h, i) shows the scattered intensity restricted to this circle and shifted such that the incident wavevector is placed at the origin. This representation directly expresses the scattering intensity as a function of the scattering vector $\vb{q}=\vb{k}_s-\vb{k}_i$. Intensities are normalized by the incident peak intensity and displayed on a logarithmic scale (dB).
}
    \label{fig:FQC TFs}
\end{figure}

The experimental structure factor is reconstructed using an Ewald-ring-based procedure. We first compute the spatial Fourier transforms of the measured complex wave fields, shown in the top row of Fig.~\ref{fig:fields}. Because the finite observation window introduces edge artifacts in Fourier space, a two-dimensional Hann window is applied to the real-space data prior to Fourier transformation. This apodization suppresses spectral leakage and significantly improves the visibility of the scattered peaks, as illustrated in the middle row of Fig.~\ref{fig:FQC TFs}.
The strongest spectral peak is identified as the incident wavevector $\vb{k}_i$. For each measurement, an annulus of width \unit{0.1}{cm^{-1}} centered at~$|\vb{k}_i|$ is extracted, thereby isolating the elastic scattering contribution satisfying $|\vb{k}_s|=|\vb{k}_i|$. The scattered intensity is then expressed as a function of the momentum transfer $\vb{q}=\vb{k}_s-\vb{k}_i$ by translating the annulus by $-\vb{k}_i$ in Fourier space. All intensities are normalized by the incident peak intensity to compensate for variations in excitation amplitude. Examples of the resulting shifted rings are shown in the bottom row of Fig.~\ref{fig:FQC TFs}.

\begin{figure}[h!]
    \centering
    \includegraphics[width=0.7\linewidth]{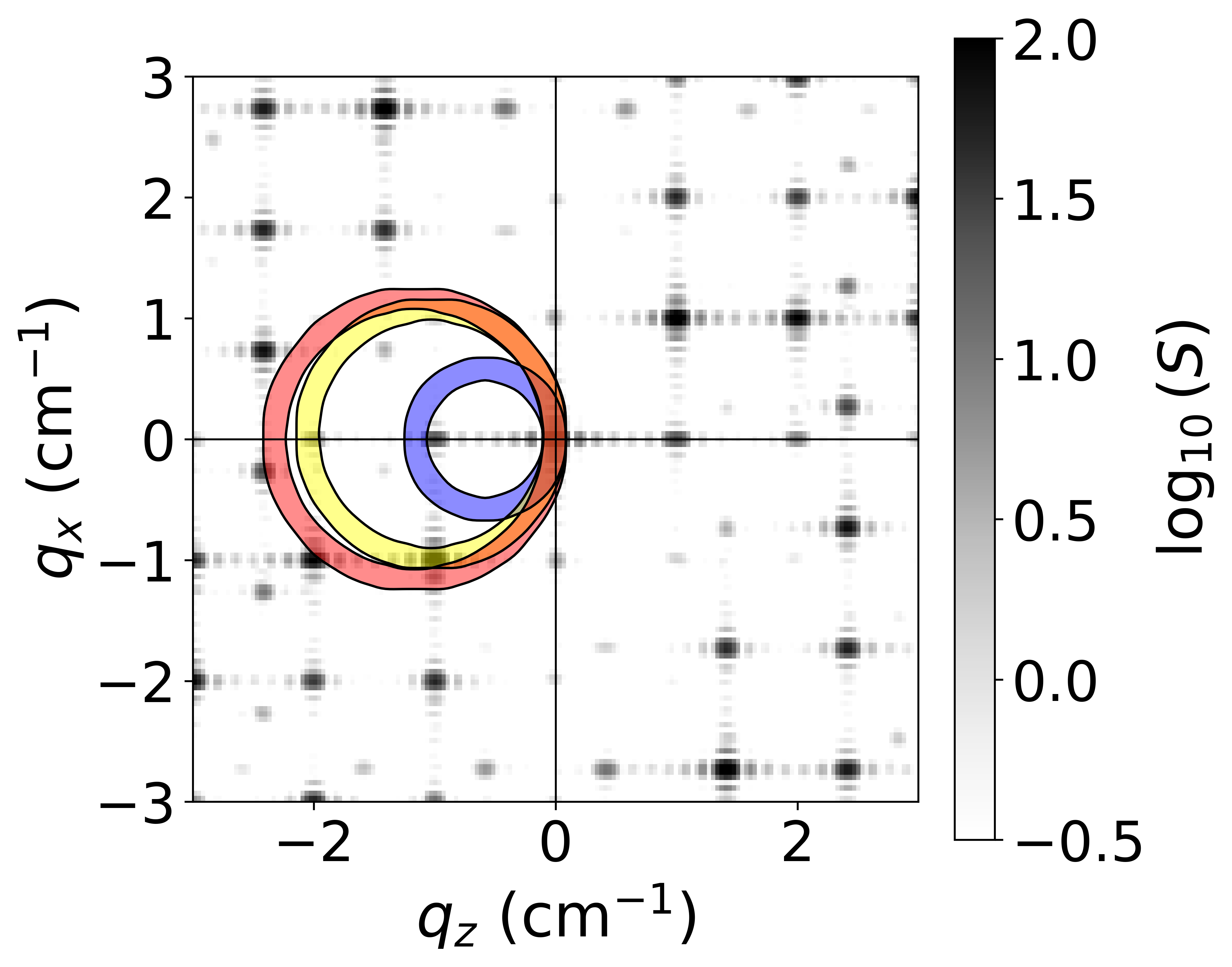}
\caption{
\textbf{Theoretical structure factor $S(\vb{q})$ of the Fourier quasicrystal} (log scale). The colored circles indicate the elastic scattering vectors accessible at three excitation frequencies: low frequency \unit{4.1}{Hz} (blue), intermediate frequency \unit{5.4}{Hz} (yellow), and high frequency \unit{6.2}{Hz} (red). These circles correspond to the set of momentum transfers allowed by elastic scattering. At low frequency, the scattering circle lies entirely within the central region where the structure factor vanishes, resulting in a suppression of scattering. As the frequency increases, the circles progressively intersect the first Bragg peaks of the structure factor, opening discrete scattering channels. This construction provides a simple geometric interpretation of the experimentally observed transition from a transparent regime to a scattering regime: scattering becomes possible as soon as the accessible scattering vectors overlap nonzero Fourier components of the structure.
}
    \label{fig:FQC TF et s}
\end{figure}

The geometrical interpretation of the reconstruction is illustrated in Fig.~\ref{fig:FQC TF et s}, where the elastic Ewald rings corresponding to three representative frequencies are superimposed on the theoretical structure factor. At low frequency, the accessible scattering vectors lie entirely within the central region where the structure factor vanishes, resulting in negligible scattering. As the frequency increases, the Ewald rings progressively intersect the first diffraction peaks, opening discrete scattering channels. At still higher frequencies, several peaks become accessible simultaneously, leading to increasingly complex scattering patterns. These observations are consistent with the transition from transparent to scattering regimes observed in the measured wave fields.

\begin{figure}[h!]
    \centering
    \vspace*{2mm}
    \includegraphics[width=0.9\linewidth]{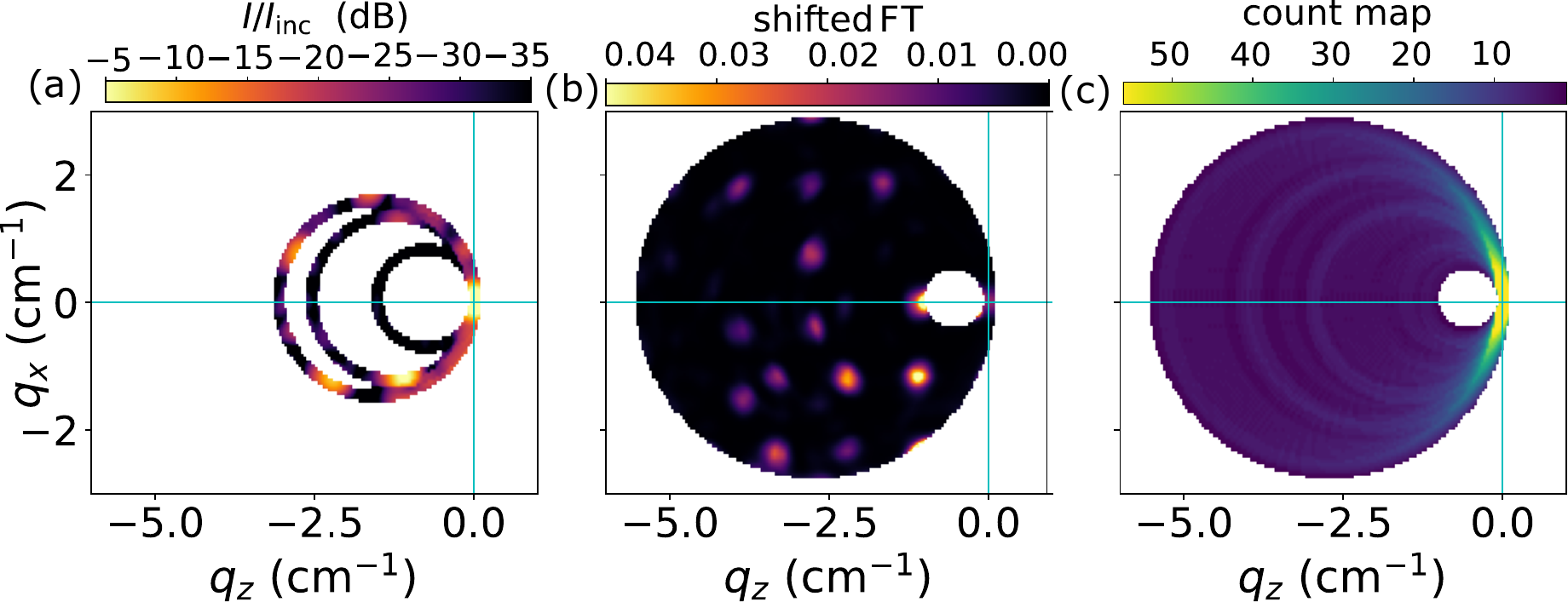}
\caption{
\textbf{Reconstruction of the structure factor by stacking elastic scattering rings measured at different excitation frequencies}.
(a)~Normalized Fourier-space intensity, divided by the incident peak intensity and displayed on a logarithmic scale, for three representative frequencies: \unit{4.1}{Hz}, \unit{5.4}{Hz}, and \unit{6.2}{Hz}.
(b)~Reconstructed structure factor obtained by stacking the shifted scattering rings measured from \unit{3.5}{Hz} to \unit{9}{Hz} in steps of \unit{0.1}{Hz}. The resulting map is shown on a linear scale after Gaussian smoothing.
(c)~Number of contributions per pixel accumulated during the stacking procedure. This map is used to normalize the intensity shown in panel (b) and reflects the reciprocal-space regions sampled by the accessible elastic scattering vectors.
}
    \label{fig:FQC_stack}
\end{figure}

In practice, the reconstruction is performed over the entire experimental frequency range. Each frequency provides information along a single Ewald ring in reciprocal space. By combining all frequencies, progressively larger portions of the structure factor become accessible. The resulting reconstruction is shown in Fig.~\ref{fig:FQC_stack}. Figure~\ref{fig:FQC_stack}a displays three representative shifted rings, while Fig.~\ref{fig:FQC_stack}b shows the structure factor obtained by stacking all measurements between 3.5 and \unit{9}{Hz}. To improve readability, the reconstructed map is convolved with a Gaussian kernel of width $\sigma=0.8$. The dominant diffraction peaks of the Fourier quasicrystal clearly emerge from the accumulated data.

Because different regions of reciprocal space are sampled with different frequencies, the number of contributions varies across the reconstructed map. The corresponding contribution count is shown in Fig.~\ref{fig:FQC_stack}c. Regions near the origin are intersected by many Ewald rings and therefore receive more contributions than outer regions. To remove this sampling bias, the accumulated intensity is divided pixel by pixel by the corresponding contribution count, ensuring a homogeneous reconstruction independently of the local sampling density.

The reconstruction described above is first performed for a single incidence direction. Experimentally, however, measurements are repeated for four distinct propagation directions. Each incidence direction probes a different region of reciprocal space, and combining all four datasets significantly enlarges the accessible reciprocal-space domain.

\bibliography{references}

\end{document}